\documentclass[prl,twocolumn,showpacs,superscriptaddress]{revtex4}
\usepackage{amsmath}    
\usepackage{amsfonts}
\usepackage{graphicx}
\usepackage{bm}
\usepackage{enumerate}
\usepackage{color}

\begin{document}
\title{Large Cross-Phase Modulation between Slow Co-propagating  
        Weak Pulses in $^{87}$Rb}
\author{Zeng-Bin Wang}
\author{Karl-Peter Marzlin}
\affiliation{Institute for Quantum Information Science,
University of Calgary, Calgary, Alberta T2N 1N4, Canada}
\author{Barry C.~Sanders}
\affiliation{Institute for Quantum Information Science,
University of Calgary, Calgary, Alberta T2N 1N4, Canada}
\affiliation{Centre for Quantum Computer Technology, Macquarie University,  
        Sydney, New South Wales 2109, Australia}

\date{\today}    

\begin{abstract}
We propose a scheme to generate 
double electromagnetically induced transparency (DEIT) and optimal
cross phase modulation (XPM) for two slow, co-propagating pulses with  
matched group velocities in a single species of atom, namely $^{87}$Rb.
A single pump laser is employed and a homogeneous magnetic field is
utilized to avoid cancelation effects through the nonlinear Zeeman effect. 
We suggest a feasible preparational procedure for the atomic initial
state to achieve matched group velocities for both signal fields.
\end{abstract}

\pacs{42.65.-k, 42.50.Gy, 03.67.-a}

\maketitle
%%%%%%%%%%%%%%%%%%%%%%%%%%%%%
\emph{Introduction:---}
The optical Kerr effect, $n = n^{(0)} +
n^{(2)} I$ (for $n$ the total refractive index, $n^{(0)}$ the linear
refractive index, $I$ the field intensity, and~$n^{(2)}$ the
optical Kerr coefficient) is invaluable for spectral
broadening  and self-focusing of laser pulses~\cite{shen}.
Large nonlinear interactions have also been used 
to generate single photons~\cite{Balic05}, 
enhanced refractive index~\cite{Zibrov96}, 
self-phase modulation~\cite{Wang01}, and Stokes and 
anti-Stokes generation~\cite{Kash99,Zibrov99}.
In cross-phase modulation (XPM), $I$ is the intensity of the other
field; thus XPM
enables the phase of each field to be controlled by the strength of
the other, which is critical for applications such as
deterministic optical quantum computation~\cite{Chu95}
and all-optical switching \cite{Rot93}.
Unfortunately $n^{(2)}$ is extremely small but can be effectively increased
via electromagnetically induced transparency (EIT)
\cite{Harris97,Lukin97,Bajcsy03},  
which yields a large optical 
nonlinearity~\cite{Schmidt96,Harris99,Hau99,Kang03}.   
Whereas these experiments demonstrated the promise of a large
nonlinear coupling for cw fields, applicability is severely
limited for propagating fields due to the different group
velocities of the two pulses, which significantly reduces the
interaction time.

Double EIT (DEIT) generalizes EIT for simultaneous action on two separate fields,
and, for slow group velocities, can effect a large XPM~\cite{Lukin00,Pet02,Matsko03,Pet04};
however DEIT has not yet been successfully realized.  
In this letter we devise a method for achieving large XPM for two slow
co-propagating pulses with matched group velocities in a single species of
atom, namely using the D1 line of $^{87}$Rb, by avoiding cancelation of nonlinearities
near resonance and decoherence due to coupling bright states, plus employing  
population transfer to achieve matched group velocities for the two pulses.

\begin{figure}
\centerline{\includegraphics[width=8.5cm]{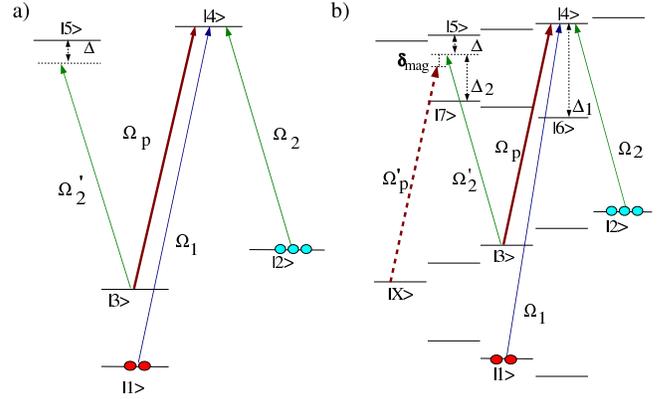}}
\caption{\label{schemeFig} Atomic level configuration for creating DEIT  
and large XPM between two    
weak signal fields 1 ($\Omega_1$) and 2
($\Omega_2$,$\Omega_2'$).
(a)~Simplified level configuration of the scheme.  
(b)~Realization of the scheme using the D1 Line of $^{87}$Rb.   
}
\end{figure}

%%%%%%%%%%%%%%%%%%%%%%%%%%%
\emph{Scheme:---}
Our scheme employs five atomic levels as shown in
Fig.~\ref{schemeFig}(a) and combines the advantageous properties of the
N-type scheme \cite{Schmidt96}, the tripod scheme \cite{Pet04}
and the M-type scheme \cite{Matsko03}.
A single pump field drives the $|3\rangle\leftrightarrow|4\rangle$ transition
with Rabi frequency~$\Omega_\text{p}$. If state $|2 \rangle $ is omitted
our scheme reduces to the N-type scheme: the AC Stark shift
created by off-resonant coupling to state $|5 \rangle$
creates a giant nonlinearity
between signal fields~1 and~2, but EIT is only realized for signal
field~1. Consequently, due to the mismatch in the group velocities, 
the interaction time is greatly reduced.  
On the other hand, if state $|5 \rangle $ is omitted we have a tripod scheme 
in which DEIT
can be achieved but for which the nonlinearity disappears at exact
two-photon resonance (as it does in the M-type scheme).
The slight detuning that is therefore necessary to create nonlinear effects  
\cite{Matsko03,Pet04} also generates linear absorption, and for a given
desired absorption rate the generated nonlinearity is suppressed by one to
two orders of magnitude.

We will show that the scheme of Fig.~\ref{schemeFig}(a)
avoids the problems associated with the schemes discussed above. 
In addition to the pump field, two signal fields with slowly varying 
field amplitude  
$\mathcal{E}_i \; (i=1,2)$ resonantly couple the
$|1\rangle\leftrightarrow|4 \rangle $ 
and $|2\rangle\leftrightarrow|4\rangle$ transitions
with Rabi frequencies $\Omega_i\equiv -|\mathbf{d}_{4i}| \mathcal{E}_i/\hbar$,
with $\mathbf{d}_{ij} = \langle i | {\bf d} | j \rangle $    
the matrix elements of the dipole moment operator~$\mathbf{d}$. In addition,
field~2 also off-resonantly couples the $|3\rangle\leftrightarrow|5\rangle$  
with Rabi frequency $\Omega'_2\equiv -|\mathbf{d}_{53}| \mathcal{E}_2/\hbar$.
If the atoms are initially prepared in a mixture \cite{avoid}
of state $|1 \rangle$ and $|2\rangle$ with density matrix
$\rho_\text{mix} = p_1 |1 \rangle \langle 1| + p_2  
|2 \rangle \langle 2|$, the two $\Lambda$-subsystems  
$|1 \rangle \leftrightarrow |4 \rangle  \leftrightarrow |3 \rangle $ and
$|2 \rangle \leftrightarrow |4 \rangle  \leftrightarrow |3 \rangle $  
induce EIT for both fields~1 and~2, i.e., DEIT.
The real part of the corresponding index of refraction $n_i$ for  
signal field $i$ then has the approximate form~\cite{Lukin97}
\begin{equation}
  n_i = 1+\eta_i\delta_{\text{p}i} \; , \;
 \eta_i \equiv \bar{\rho} \, p_i \, |\mathbf{d}_{4i}|^2
    /(2\hbar\varepsilon_0 |\Omega_\text{p}|^2)
\label{refindSimple}
\end{equation}    
for $i=1,2$, $\delta_{ij}\equiv\delta_j-\delta_i$, $\bar{\rho}$ the atomic number
density, and $p_i$ the initial population in state $|i \rangle$.
For~$E_{ij}\equiv E_j-E_i$, the detuning of the signal field frequency~$\omega_i$    
and the laser pump frequency~$\omega_\text{p}$ are~$\delta_i=\omega_i -E_{i4}/\hbar$    
and $\delta_\text{p}=\omega_\text{p} -E_{34}/\hbar$, respectively.
Because Eq.~(\ref{refindSimple}) varies strongly with the  
two-photon detuning $\delta_{\text{p}i}$ the group velocity is very small.  
Below we will show how the group velocities can be made equal by
manipulating the populations~$p_i$.

The off-resonant  
coupling for transition $|5\rangle\leftrightarrow|3\rangle$ caused by field~2
modifies the tripod scheme by
producing an AC Stark shift for state~$|3\rangle$ given by    
\begin{equation}
        \Delta E_3 =\hbar |\Omega_2'|^2/\Delta,\quad\Delta = \omega_2 -E_{35}/\hbar \;          
\label{acstark}
\end{equation}    
with $\Delta$ the (large)  
detuning of field~2 with respect to the    
$|3 \rangle\leftrightarrow |5 \rangle$ transition.
This energy shift implies that the
pump frequency $\omega_\text{p}$ is somewhat detuned
from the $|3 \rangle\leftrightarrow|4 \rangle $ transition;
hence the index of refraction for this EIT medium changes due to the
dispersion relation of Eq.~(\ref{refindSimple}).
  
Replacing $\delta_{\text{p}i}$ by $\delta_{\text{p}i}-\Delta E_3/\hbar$    
and inserting this into Eq.~(\ref{refindSimple}) yields
\begin{equation}
 n_1  = 1 +\eta_1(\delta_{\text{p}1}-\chi I_2),\    
 \chi =|\textbf{d}_{53}/\hbar|^2/{c\varepsilon_0\Delta}
 \label{refindSimpleNL}
\end{equation}    
with intensity $I_i = | \mu_0^{-1} {\bf E}_i\times {\bf B}_i^*|=  
c \varepsilon_0 |{\cal E}_i|^2$  
corresponding to the modulus of the complex Poynting vector.
  
We find that this intuitively appealing phenomenological  
derivation agrees with a rigorous calculation based on third-order
time-dependent perturbation theory in the weak signal fields. Neglecting all
decoherence effects and assuming a resonant pump field ($\delta_\text{p}=0$), the
Schr\"odinger equation yields the index of refraction  
\begin{align}    
  n_1 &= 1+\eta_1(\delta_1-\chi\,I_2)
\label{simpSol} \\
  n_2 &= 1+\eta_2(\delta_2-\chi\,I_2) -\eta_1\chi\,I_1  
\nonumber
\end{align}    
with the self-phase modulation (SPM) coefficient for field~2 given by
$\eta_2\chi$ and cross-phase modulation (XPM) coefficient for both fields given by
$\eta_1\chi$. Both XPM and SPM coefficients are given to first order in  
$1/\Delta$ and to second order in $1/|\Omega_\text{p}|^2$.
A more detailed analysis (see below) shows that this simple result is  
qualitatively correct if the detuning $\Delta$ is much larger than the decay
rate, and the AC Stark shift~(\ref{acstark}) is smaller than the width of the EIT
transparency window. As this scheme shares the advantages and avoids the
disadvantages of the N-type, tripod and M-type schemes, it yields the optimal
XPM based on EIT techniques.  

%%%%%%%%%%%%
\emph{Implementation with $^{87}$Rb:---}
Our scheme can be realized in a gas with a single species of atoms; a specific
implementation using the D1 line of $^{87}$Rb is illustrated in  
Fig.~\ref{schemeFig}(b). A homogeneous magnetic field parallel
to the laser propagation minimizes coupling to states
that are not part of the scheme in Fig.~\ref{schemeFig}(a).
It is of particular importance to break  
two-photon resonance for the $\Lambda$ subsystem  
$|3 \rangle \leftrightarrow |5 \rangle  \leftrightarrow |\text{X} \rangle $.
If this is not the case a further EIT transition that transfers
atoms from state $|3 \rangle$ to state  
$|\text{X} \rangle$; this would remove the crucial energy shift
(\ref{acstark}) and hence destroy XPM.
The linear Zeeman effect is not suitable to break two-photon resonance.
However, for large enough magnetic fields the Zeeman splitting depends
nonlinearly on the magnetic quantum number without affecting the
corresponding selection rule \cite{Sarkisyan03}.
Numerical diagonalization of the Hamiltonian for a magnetic field
of $B=150$~G shows that the energy
differences $E_{32}$ and $E_{\text{X}3}$  
differ by an amount $\delta_\text{mag}=(E_{32}-E_{\text{X}3})/\hbar=-12.9$~MHz.
This is much larger than the EIT transparency window~\cite{Schmidt96}
$\delta\tau=\text{min}\lbrace\Omega_\text{p}',|\Omega_\text{p}'|^2/\gamma\rbrace$
and breaks two-photon resonance for the
$|3 \rangle \leftrightarrow |5 \rangle  \leftrightarrow |\text{X} \rangle $
transition. We therefore can neglect state $|\text{X} \rangle $ in  
our considerations; this approximation is confirmed by our numerical
simulations (see below).

\begin{figure}[b]
\centerline{\includegraphics[width=8cm]{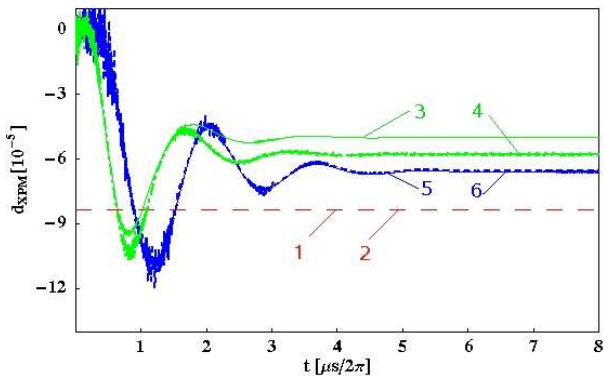}}
\caption
{\label{abXPM}  
Mean cross dipole moment $d_{\text{XPM}}$  
of Eq.~(\ref{crossdipole}) as a function of time for signal field 1 and 2.
Lines 1, 2 represent the result of SAT (Eq.~(\ref{simpSol})),
which is identical for both signal fields.
Lines 3, 4 (5, 6) represent the results of EAT and NUM for signal field 1
(2), respectively.
The parameters used for these solutions are  
$\Omega_1=0.68$~MHz,    
$\Omega_2=-0.55$~MHz,    
$\Omega_\text{p}=4.06$~MHz,    
$\Delta=-134.58$~MHz,    
$\Delta_1=894.93$~MHz,
$\Delta_2=621.85$~MHz,    
$\delta_\text{mag}=-12.93$~MHz, and
$\gamma=5.73$~MHz, $\bar{\rho}=10^{14}$~cm$^{-3}$
and $B=150$~G, with initial atomic state
$\rho_\text{mix} = 0.4|1 \rangle \langle 1| + 0.6|2 \rangle \langle 2|$.
}
\end{figure}

We analyzed this specific scheme in three ways:
the simple analytical theory (SAT) described above yields Eq.~(\ref{simpSol}) for the XPM coefficient,
a more elaborate analytical theory (EAT), and a numerical simulation
(NUM). The elaborate theory EAT is based on  
third-order time-dependent perturbation theory to solve the
Schr\"odinger equation with the evolution restricted
to states $|1 \rangle\leftrightarrow|7 \rangle$ of  
Fig.~\ref{schemeFig}(b). Spontaneous emission is included
in the non-Hermitean atomic Hamiltonian
\begin{equation}    
  H = \hbar
  \left(\begin{array}{ccccccc}
  \delta_1   &0   &0  &\Omega_1^{*}  &0 & \Omega_1'^{*} & 0 \\
  0  &\delta_2 &0 &\Omega_2^{*} &0 &\Omega_2''^{*} & 0 \\
  0  &0 &\delta_\text{p} &\Omega_\text{p}^{*} &\Omega_2'^{*}  &\Omega_\text{p}'^{*}
  &\Omega_2'''^{*}\\
  \Omega_1 & \Omega_2  &\Omega_\text{p}  &-i\frac{\gamma}{2} &0  &0 & 0 \\
  0 &0 &\Omega_2' &0 &\tilde{\Delta} &0 & 0 \\
  \Omega_1'  &\Omega_2''  &\Omega_\text{p}' &0 &0    
  &\tilde{\Delta}_1 & 0 \\
  0  &0  &\Omega_2''' &0 &0 &0 &\tilde{\Delta}_2
  \end{array}\right) \; ,
\label{Heq}
\end{equation}
with detunings (see Fig.~\ref{schemeFig}(b))  
$\Delta_1 = \omega_1 -E_{16}/\hbar$,  
$\Delta_2 = \omega_2 -E_{37}/\hbar$ and
phenomenological complex quantities  
$\tilde{\Delta}_i=-\Delta_i-i\frac{\gamma}{2}$~\cite{Harris98}.  
We obtained an accurate analytical result,
but the expression is unwieldy and not displayed here. Instead we
compare it with the numerical simulation NUM in which we solved
the time evolution of the atomic density matrix based on the
full Lindblad master equation for all 16 states of Fig.~\ref{schemeFig}(b),
including spontaneous emission based on all Clebsch-Gordan
coefficients and 6-j symbols as specified by Steck \cite{SteckRb87}.
Below we will also briefly discuss numerical results that include
dephasing and transit broadening which is important for hot atomic
gases.
XPM between the two signal beams is proportional to that part of
the atomic mean dipole moment which couples to signal field $i$ and  
vanishes if the intensity of the other signal field is zero.
To describe this we define a dimensionless mean cross dipole moment by
\begin{equation}    
  d_\text{XPM}  
  = \frac{1}{e a_0} \left (
   \left |\text{Tr}(\rho \mathbf{d}) \right |_{\mathcal{E}_{3-i} \neq 0}  
  -\left |\text{Tr}(\rho \mathbf{d}) \right |_{\mathcal{E}_{3-i} = 0}
  \right ) \; ,
\label{crossdipole}
\end{equation}  
with $e$ the electron's charge and $a_0$ the Bohr radius. $d_\text{XPM}$
is a direct measure of the mutual inteaction between the signal pulses;
in SAT and EAT it is proportional to the XPM terms, in NUM it also contains
higher order terms.
The result of NUM for $d_\text{XPM}$
are shown in (Fig.~\ref{abXPM})
and show excellent agreement with the analytical results EAT
and qualitative agreement with the simple theory SAT.
We have used Steck's spectroscopic data~\cite{SteckRb87}  
and the parameters given in the figure caption.

The XPM phase shift can be identified as the difference between
the phase factors acquired by signal field $i$  
if the second signal field is switched on and off,  
respectively, after it has been propagating for $L=1.6$ mm (double the    
Rayleigh length). This definition amounts to
\begin{equation}    
  \phi_i^\text{XPM} = \frac{\omega_i}{c} L  
  \Big ( n_i(\mathcal{E}_{3-i} \neq 0) -
  n_i(\mathcal{E}_{3-i} = 0) \Big ) \; ,
\end{equation}    
for $i=1,2$ and with the refractive index evaluated using the
result of EAT  
for the steady state value of the atomic dipole moment.
Fig.~\ref{abXPM} indicates that steady state is already
achieved after $0.7\, \mu$s.
The stationary values are given by    
$ \phi_1^\text{XPM} =1.11$, corresponding to an effective XPM coefficient  
$(\eta_1\chi)_1^\text{eff}=0.27\text{cm}^2/$W, and    
$\phi_2^\text{XPM} =0.84$, corresponding to an effective XPM coefficient  
$(\eta_1\chi)_2^\text{eff}=0.20\text{cm}^2/$W.    
The difference of XPM for field 1 and 2 is due to decoherence effects
and coupling to off-resonant states.
The pulses are attenuated by about 20\% due to off-resonant
coupling to states $|6 \rangle $ and $|7 \rangle $.  
In addition, the nonzero detuning associated with the  
energy shift (\ref{acstark}) leads to  
an intensity-dependent attenuation of about 3\%.

In summary, the three theoretical results SAT, EAT, and NUM
indicate that the five-level scheme of Fig.~\ref{schemeFig}(b)
would work well even if all 16 hyperfine states are included.
The off-resonant coupling to the additional states only
results in the small, very rapidly
oscillating perturbations displayed in Fig.~\ref{abXPM}
as an increased width of lines 4 and 6.

%%%%%%%%%%%%%%%%%
\emph{Preparation of the initial state:---}
For pulsed signal fields maximum interaction time is achieved
for DEIT with equal group velocities.  
From Eq.~(\ref{simpSol}) one can infer that
the group velocity $v_i$ of signal field $i$
is proportional to $v_i \propto p_i|\textbf{d}_{4i}|^2$.
It is therefore possible to achieve $v_1 = v_2$ by preparing suitable
populations $p_i$ of the initial density matrix
$\rho_\text{mix} = p_1 |1 \rangle \langle 1| + p_2 |2 \rangle \langle 2|$.
We propose the following procedure.
We first pump all atoms to state $|4 \rangle $ whence they decay
to states~$| 1 \rangle$ and~$|2 \rangle$ plus other states.
To maximize the EIT effect one may repump atoms in all states but
$| 1 \rangle$ and $|2 \rangle$ into state $|4 \rangle $ until    
the atomic state is well approximated by $\rho_\text{mix}$ with
$p_i = \gamma_{4i}/(\gamma_{41}+\gamma_{42}) \propto |\textbf{d}_{4i}|^2$,    
where $\gamma_{4i}$ is the decay rate from state $|4 \rangle $ to state
$|i \rangle $. With these populations the group velocities would be
proportional to  $v_i \propto |\textbf{d}_{4i}|^4$ and thus differ
significantly. We therefore induce as a last step in the atomic
state preparation a Raman transition between $|1 \rangle $ and  $|2 \rangle $.
Ideally this exchanges the populations, $p_1 \leftrightarrow p_2$, resulting  
in $p_i=|\textbf{d}_{4,3-i}|^2/({|\textbf{d}_{41}|^2|+\textbf{d}_{42}|^2})$
so that both group velocities are now proportional to
$ |\textbf{d}_{41}|^2  |\textbf{d}_{42}|^2$ and therefore equal.
We remark that the Raman transition is not required to work perfectly.
An exchange of only 90\% of the populations would lead to
a difference of 10\% in the group velocities, for instance.

%%%%%%%%%%%%%%%%%%%%%%%%%%%%
\emph{Maximum phase shift for pulses at single-photon level:---}
One prominent application of XPM  would be
the creation of a controlled phase gate for photonic qubits  
\cite{Nielsen}.
We therefore estimate the maximal XPM phase shift
achievable for two Gaussian pulses at single-photon level propagating
through DEIT media. To do so we generalize the expression of XPM Kerr coefficient (\ref{simpSol})
by replacing $\eta_1$ by the standard EIT expression  
that includes spontaneous emission ~\cite{Harris97, Lukin97}  
for $\Lambda$ atoms in a resonant pump beam,
$\tilde{\eta}_1 = \eta_1 |\Omega_\text{p}|^2/
  (|\Omega_\text{p}|^2 -\delta_1 (\delta_1+i \gamma/2))$.
To first order in the detuning $\delta_1$ we have
$\tilde{\eta}_1 \approx \eta_1 + i \tau \delta_1$
with $\tau \equiv \eta_1 \gamma /(2 |\Omega_\text{p}|^2)$.  
Within the paraxial approximation,
the slowly varying electric field amplitude of
a Gaussian pulse  of duration $T$ and minimum
(1/e intensity) waist $w_0$ at single-photon level
in a medium that is described    
by $\tilde{\eta}_1$ propagates
according to
\begin{eqnarray}    
  {\cal E} &=&    E_\text{max}  \frac{ w_0^2 }{\Theta }\,
   \frac{T}{\Xi}
   \text{exp}\big\{
   - \frac{\scriptstyle x^2 + y^2}{\scriptstyle 2\Theta}
   - \frac{\scriptstyle \left (\frac{\scriptstyle z}{\scriptstyle v_\text{gr}} -t  \right
   )^2}{\scriptstyle \Xi^2}    
   \big\}
\\    
  E_\text{max} &=&  
  \sqrt{\frac{\hbar \omega}{\sqrt{2\pi^3}\varepsilon_0 c T w_0^2 }}
\end{eqnarray}    
with $\Theta \equiv w_0^2 + i \frac{z}{k}$ and
$\Xi \equiv \sqrt{T^2 + 4 k z \tau}$.
As for any Gaussian beam the field intensity is only close to its
maximal value over a propagation length $L = 2 z_R$, with    
$ z_R = k w_0^2$ the Rayleigh length. In addition, the EIT transparency window
attenuates the pulse through the factor $\Xi$. Requiring that $\Xi$
changes little over $z_R$ implies a minimum pulse duration of
$T = 2 \sqrt{\tau k z_R}$. Using these values we can estimate the
maximum phase shift $\phi^\text{max}$ as the product of $L $ and the XPM induced
change in the wavevector $k = n \omega/c$,
\begin{eqnarray}    
  \phi^\text{max} &=& L \frac{\omega}{c} \chi^\text{XPM} c \varepsilon_0  E_\text{max}^2
  \\
\label{MaxPhase}   
  &=& -\frac{3\sqrt{6}}{4\pi}\frac{\lambda}{ w_0}
  \frac{ \gamma}{\Delta} \sqrt{\bar{\rho} k^{-3}}
  \approx 0.58
  \frac{\lambda}{ w_0}
  \frac{ \gamma}{\Delta} \sqrt{\bar{\rho} k^{-3}}
\nonumber
\end{eqnarray}    
To avoid decoherence we need $\gamma /\Delta \ll 1$. 
The diffraction limit~\cite{Pet02} implies $\lambda / w_0 < 1$ and 
we estimate that resonant
dipole-dipole interaction would modify $\phi^\text{max}$
for $\bar{\rho} k^{-3} \simeq 1$. Thus, the maximum XPM phase shift between two
photons would in this approach be of the order of 0.1 rad.
This is considerably smaller than previous estimates of more than one rad
\cite{Harris99,Lukin00} but may be overcome by combining our scheme with 
the ideas presented in Ref.~\cite{Andre05}.
The difference arrives from imposing different
constraints (dipole-dipole interaction) and our neglect of  
ground-state decoherence,
which is reasonable for trapped ultracold atoms.

%%%%%%%%%%%%%%%%%%%%%%%%%
\emph{Limitations:---}
The numerical example presented above is based on the parameters for a dense
ultracold gas such as a trapped elongated Bose-Einstein condensate.
A mutual phase shift of the order of $\pi$ could then be obtained
for pulses containing hundreds of photons.
In a hot atomic gas (e.g., T=440K and $\bar{\rho} \sim    
10^{14} \text{cm}^{-3}$),    
the efficiency of this scheme is mainly limited by the atomic motion.    
Although the Doppler effect is canceled in the two-photon detuning
$\delta_i -\delta_\text{p}$ for co-propagating light fields,
the atomic motion narrows the transparency window according to
$\delta_{\tau}\leq \Omega_\text{p}^{2}/\Delta_{D}$~\cite{Kash99, Ye2002}, where    
$\Delta_{D}\approx 500$~MHz  is the mean Doppler detuning.  
The XPM phase shift would then be two orders of magnitude
smaller than the optimal value.
Transit broadening and ground state dephasing \cite{Beau04} 
due to the atomic motion in and out of the
laser fields and collisions between atoms are also detrimental.
For laser pulses with a width of 2mm in a hot atom setup using buffer gas \cite{Zibrov99},
rates less than 1kHz are possible. 
Numerical simulations indicate that our scheme would produce 
$0.4$rad XPM (with 40\% attenuation) 
for two weak pulses with $10^2$ photons,
a transit broadening rate of $1$kHz, and a ground state 
dephasing rate of $1$kHz. 

%%%%%%%%%%%%%%%

\emph{Conclusion:---}
Cross-phase modulation between weak signal fields is one of the most
significant challenges for nonlinear optical switching and quantum
information. We have proposed an approach to achieve strong
cross-phase modulation under realistic conditions for $^{87}$Rb
in which matched group velocities for two interacting pulses 
can be realized.
Our approach incorporates all of the following
desirable conditions: co-propagating laser beams to avoid Doppler shifts, 
a single atomic species, and only a single pump field. 
We show that our scheme yields optimal XPM for pump schemes based on DEIT,
and we estimate the maximum phase shift to be 0.1 rad for Gaussian  
pulses at single-photon level.
To avoid decoherence effects Bose-condensed gases would be preferred, 
but we expect the scheme also to be effective for
thermal gases if sufficiently strong signal pulses are used.

%%%%%%%%%%%%%%%%
\emph{Acknowledgments:---}
We thank A.\ I.\ Lvovsky, J.\ Appel, F.\ Vewinger, and A.~Steinberg    
for helpful discussions and acknowledge support from iCORE, NSERC, CIAR, and ARC.


\begin{thebibliography}{99}
\bibitem {avoid} A coherent superposition of the two states should be
        avoided because it may correspond to the dark state of the    
        $\Lambda$ system $| 1 \rangle \leftrightarrow |3 \rangle \leftrightarrow |2 \rangle$.
        In this case the signal fields would not be coupled to the atoms at all.
\end{thebibliography}
\end{document}